\documentclass{article} 
\usepackage{arxiv_style}
\usepackage{times}
\usepackage{graphicx}
\usepackage{amssymb}
\usepackage{amsmath}
\usepackage{url,hyperref}
\usepackage[numbers]{natbib}
\usepackage{physics}
\usepackage{qcircuit}
\usepackage{subfig}
\usepackage{authblk}
\usepackage{multicol}
\usepackage{authblk}
\usepackage{float}
\usepackage{xcolor}
\usepackage{algorithm}
\usepackage{algpseudocode}
\usepackage{enumitem}
\usepackage{stmaryrd}

\makeatletter

\let\@fnsymbol\@arabic
\makeatother

\begin{document}

\title{Monte Carlo particle transport on quantum computers}

\author{No\'e Olivier}
\author{Michel Nowak}

\affil{\textit{Thales Research and Technology, Palaiseau, France}}

\maketitle
\thispagestyle{empty}
\let\thefootnote\relax\footnotetext{noe.olivier@thalesgroup.com}
\let\thefootnote\relax\footnotetext{michel.nowak@thalesgroup.com}

\begin{abstract}
    Monte Carlo particle transport codes are well established on classical hardware and are considered as the reference tool for nuclear applications.
    In a growing number of domains, the design of algorithms is progressively shifting towards the field of quantum computing, where theoretical speedups over their classical counterparts are expected.
    In some of these domains, Monte Carlo methods have already been converted to a quantum computing friendly setup where the expected and observed gain in complexity is quadratic.
    Surprisingly, particle transport has been left aside along the path and lacks an implementation on these new architectures.
    In this work, a numerical scheme for particle transport based on discrete-time quantum walks and its combination with the amplitude amplification routine are proposed to catch up with the promised speedups.
\end{abstract}

\begin{multicols}{2}

\section{Introduction}
\label{sec:introduction}
Monte Carlo particle transport codes~\cite{brun2014tripoli,romano2013openmc,kulesza2022mcnp} are well established and yield with reference calculations for the field of nuclear energy, and more generally for radiation protection.
Continuous development on classical computers still succeeds at improving the convergence rate of these codes, in particular by leveraging GPU capabilities~\cite{tramm2024performance}. However running those codes on HPC clusters remains costly and resource demanding.

As quantum computers are emerging, researchers develop and implement algorithms that can leverage these architectures in a more and more efficient way.
Boosted by promising theoretical speedups, these developments are now guided by real-world applications in many fields.
Quantum friendly Monte Carlo methods are already proposed in a wide range of fields, starting from finance~\cite{rebentrost2018quantum, kaneko2021quantum}, to theoretical~\cite{montanaro2021montecarlo} and practical integration~\cite{yu2020practical, noto2020quantum}, with the motivation of a quadratic speedup.

The goal of this work is to investigate how the quantum formalism can help accelerate Monte Carlo particle transport simulations.
Our motivation relies on three observations.
First, the amplitude amplification and estimation algorithm~\cite{brassard2002quantum} is used to integrate stochastic processes on quantum computers.
Amplitude amplification has been proposed as an improvement of Grover's algorithm.
The costly part of this algorithm is being eliminated and replaced with low depth subroutines~\cite{withoutsphae2020}.
Second, amplitude amplification and estimation has been proved to be efficient on real hardware~\cite{giurgica2021low,certo2022benchmarking}, even in a Monte Carlo framework on photonic hardware~\cite{rebentrost2018photonic}.

The goal of the work is to explore the possibility of computing particle transport on quantum computers.
Section~\ref{sec:classical_approach} briefly introduces the Monte Carlo method on classical computers.
Section~\ref{sec:quantum_appoach} gathers our proposal for a numerical scheme on quantum computers.
Section~\ref{sec:numerical_experiments} presents the numerical experiments that we conducted after formulating the circuit.

\section{Classical approach}
\label{sec:classical_approach}
Consider $\pi$, a probability distribution from which we can
sample trajectories, that carries information about the studied system. Note $\mathbb{T}$ the space in which
the trajectories belong.
A batch of Monte Carlo sampling consists in generating a set of trajectories
$S\in\mathbb{T}^N$ where $N$ is the number of trajectories sampled.
The procedure to sample such trajectories is to follow the next steps

\subsubsection*{Source sampling}
The trajectories are initialized at source points~\cite{coste2013neutronique}.
The probability density with which this happens is thus
\begin{equation}
   \pi_\text{source}(x) = \frac{S(x)}{\int{S(x')dx'}}.
\end{equation}

\subsubsection*{Flight length sampling}
Once the trajectories have been initialized, the particles experiment an exponential flight, that is sampled with the following probability distribution
\begin{equation}
   \pi_\text{flight}(\rho) = \Sigma_t e^{\Sigma_t \rho}\text,
\end{equation}
where $\rho$ is the flight length and $\Sigma_t$ the total cross section of the material at which the particle has been sampled from a source or a previous collision.

\subsubsection*{Scattering sampling}
The scattering is made of two steps. The first step consists in choosing the isotope on which the particle collides.
The second step is to compute the outgoing angle of scattering.
In order to simplify our developments, we consider that the scattering is isotropic and that the materials are homogenized into macroscopic cross sections.
The probability of scattering would thus be
\begin{equation}
    p_s=\frac{\Sigma_s}{\Sigma_{t}}\text.
\end{equation}
\subsubsection*{Iterating}
The flight and collision steps are repeated until the particle is absorbed by the material.

\subsubsection*{Scoring a response}
Once these trajectories are sampled, one can integrate a response $\varphi$ over those trajectories.
\begin{equation}
   R=\frac{1}{N}\sum\limits_{\mathcal{T}\in S}\varphi(\mathcal{T})\text,
\end{equation}
or focus on the trajectories themselves.

\subsection{Chosen setup}
\begin{figure}[H]
    \centering
    \includegraphics[width=0.95\linewidth]{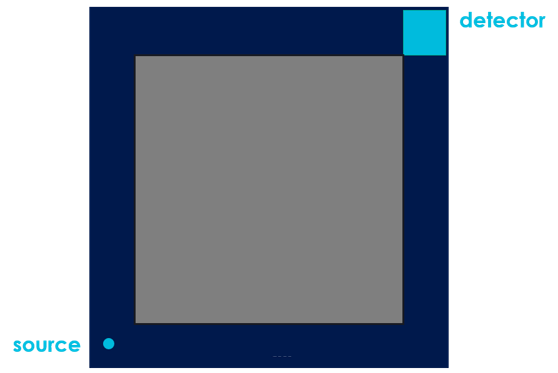}
    \caption{Bypass geometry.}
    \label{fig:bypass_geometry}
\end{figure}

For our numerical experiments, we'll consider the geometry depicted in Figure~\ref{fig:bypass_geometry}.
Both materials have $\Sigma_t = 1$.
The obstacle (in gray) has $\Sigma_s=0.1$.
The arms (in blue) has $\Sigma_s=0.9$.

\section{Quantum approach}
\label{sec:quantum_appoach}
In this section, we propose an extension of particle transport on classical computers to the field of quantum computing.
This extension comes with various simplifications and intends to model as faithfully as possible the dynamics of the transport with quantum computing tools.
There are many types of responses that one might want to score. Here, we tackle the problem of shielding calculations, where a source of particle is expected to collide in a detector domain.

First of all, the core of a particle transport code relies on frenetic calls to the geometry.
By geometry we mean the position of the materials, the boundary conditions and the composition of these materials.
The probabilities of interaction will vary depending on the position of the particle tracked, as well as on the isotopes present at this location. These probability tables are significantly demanding in memory.

In this paper, the proposed quantum approach relies on the formalism of discrete-time quantum walks.
This formalism treats the position of the particle as being a wave function defined on a grid.
Upon iterating on a coined controlled shift operator, the trajectory evolves until it is measured and the positions can be retrieved in a classical register.

\subsection{Trajectory encoding}

\begin{figure*}[ht]
    \centering
    \subfloat[10x10 nodes 7 qubits]{\includegraphics[width=0.3\linewidth]{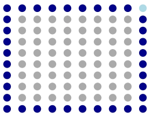}}
    \subfloat[20x20 nodes 9 qubits]{\includegraphics[width=0.3\linewidth]{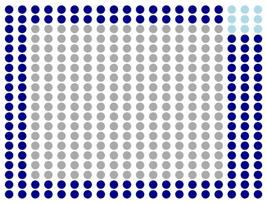}}
    \subfloat[80x80 nodes 13 qubits]{\includegraphics[width=0.3\linewidth]{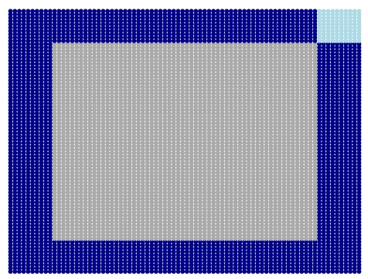}}
    \caption{Discretized geometry with various number of nodes per axis.}
    \label{fig:bypass_discretized_geometry}
\end{figure*}

Let's consider the problem of two-dimensional particle transport in a heterogeneous medium.
In order to encode the trajectories in a quantum setup, a first natural choice is to discretize the geometry on a graph. Typically, our algorithm scans the geometry studied and discretizes it in a two-dimensional grid whose cells correspond to nodes of the graph.
Therefore, we define a \textit{position} quantum register of $n_\text{pos} = n_x + n_y$ qubits, allowing the state representation of all the cells in a grid of dimensions $2^{n_x} \times 2^{n_y}$ as depicted in Fig \ref{fig:bypass_discretized_geometry}.
The quantum state corresponding to a single cell is described by
\[
\ket{q_{n_\text{pos}-1} \ldots q_1, q_0} = \ket{x_{n_x-1} \ldots x_1 x_0 y_{n_y-1} \ldots y_1 y_0}
\]
By convention the position of the particle on the horizontal (resp. vertical) dimension of the grid is given by the position qubits $\ket{x_{n_x-1} \ldots x_1 x_0}$ (resp. $\ket{y_{n_y-1} \ldots y_1 y_0}$).
For simplicity, we consider that $n_x = n_y$ even though the proposed algorithm allows a more general setting.

In order to encode trajectories of multiple collisions, we rely on the quantum walk formalism, which tracks a particle after N-collisions~\cite{kempe2003quantum, szegedy2004quantum, lovett2010universal}.

\subsection{Source}

We call source the set9 of nodes in the graph from which the particles are generated.
In other words, the first step of the algorithm is to prepare a state in the position register that corresponds to a uniform superposition over the source nodes. This corresponds to the initialization of the particle trajectories.

For our purpose, let's consider a punctual source as any other sources can be expressed as a sum of punctual ones and can be implemented using the method developed by~\cite{mottonen2004transformation}.
The initial state $\ket{\psi_\text{source}}$ is prepared such that it corresponds to the punctual source node by applying single-qubit Pauli X gates on a selection of qubits from the position register.

\subsection{Transport}

The modeling of the diffusion of the particles across the geometry is done via the definition of a discrete-time quantum walk.
The time evolution of the process is discretized in a number of iterations, each of which can be described by the following operations:
\begin{itemize}[label=$-$, leftmargin=1cm, parsep=0cm, itemsep=0cm, topsep=0cm]
    \item \textbf{Coin operator}: it defines the direction of propagation.
    \item \textbf{Boundary conditions}: it defines the propagation conditions on the border of the geometry.
    \item \textbf{Shift operator}: it changes the position of the particle considered based on the direction selected.
\end{itemize}

\subsubsection*{Position dependent coin operator}

\begin{figure}[H]
    \centering
    \includegraphics[width=0.95\linewidth]{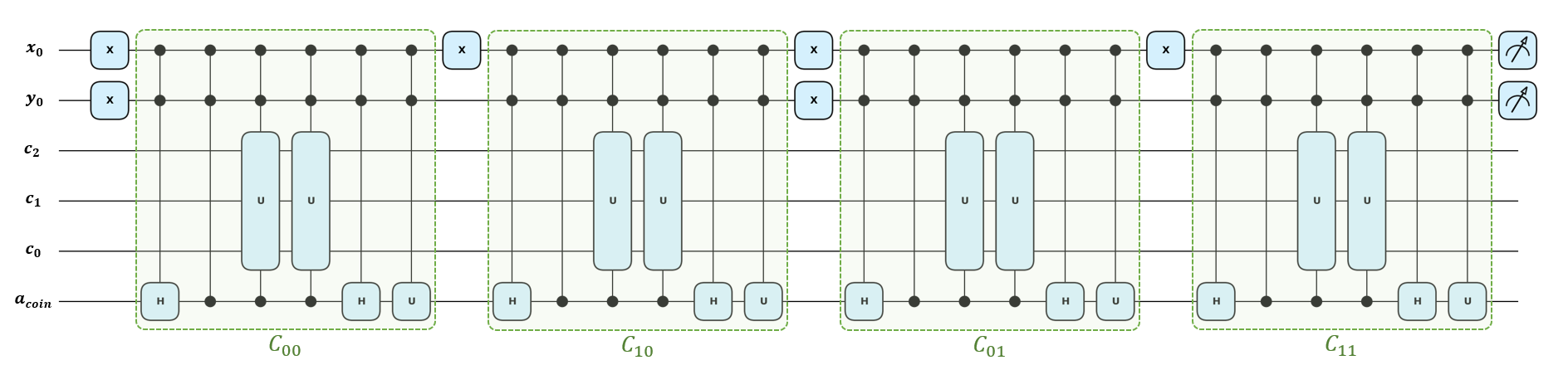}
    \caption{Circuit for the position-dependent coin operator.}
    \label{fig:position_dependent_coin}
\end{figure}

The coin operator acts on a specific \textit{coin} register and results in a state describing a direction of propagation.
Most works on coin-based quantum walks consider global coins, such as the Hadamard coin and the Grover coin, which act on the coin register independently from the state of the qubits in the position register.
However, such coins are not well adapted to the modeling of particle transport.
Here, the probabilities of interaction with the geometry are local: the medium of propagation is generally not homogeneous and the behavior of the particles depends strongly on their location.
Therefore, it is necessary to rely on a local coin operator in order to transport the particles according to the local cross sections.
We propose to rely on the "naive" position-dependent coin operator introduced by~\cite{nzongani2023quantum} for the overall structure of the circuit. As shown in Fig~\ref{fig:position_dependent_coin}, the depth of the circuit increases exponentially with the number of position qubits since one coin operator must be defined for each of the possible position states. This paper also presents another implementation of such local coin operator which allows a linear-depth circuit at the cost of an exponential number of ancillary qubits.

In our algorithm, the coin register is composed of three qubits to account for five possible movements: right $\ket{\rightarrow}$, up $\ket{\uparrow}$, left $\ket{\leftarrow}$, down $\ket{\downarrow}$ and staying at current location $\ket{\circlearrowright}$.
Using the convention $\ket{c_2 c_1 c_0}$, the following Table describes the implementation of these possibilities.
\begin{center}
    \begin{tabular}{|c|c|}
        \hline
        Coin qubits & Directions \\
        $\ket{c_2 c_1 c_0}$ & \\ [0.3ex]
        \hline\hline
        $\ket{100}$ & $\ket{\rightarrow}$\\
        $\ket{110}$ & $\ket{\uparrow}$\\
        $\ket{101}$ & $\ket{\leftarrow}$\\
        $\ket{111}$ & $\ket{\downarrow}$\\
        $\ket{0\cdot\cdot}$ & $\ket{\circlearrowright}$\\ [0.3ex]
        \hline
    \end{tabular}
\end{center}

The position-dependent coin operator is controlled by the qubits of the position register and associates different probabilities for the five possible coin results depending on the geometry.
The challenge consists in designing a coin operator that takes into account the different interactions with the material.
As a first simplification, we focus on macroscopic cross sections, such that we average the cross sections locally over all isotopes.
Let's consider both the scattering and absorption effects and their contributions to the cross sections $\Sigma_\text{s}^{x,y}$ and $\Sigma_\text{a}^{x,y}$ respectively.
Instead of killing a particle that would be subject to absorption, we choose to keep it in the same location with some probability $p_\text{a}^{x,y}$ by considering a lackadaisical quantum walk~\cite{inui2005one} and adding a self-loop of such weight. Lackadaisical quantum walks have been shown to be particularly useful in searching for a marked element in 2D-grids~\cite{nahimovs2019lackadaisical, wong2018faster}. In our model the marked element would correspond to the location of a sensor that detects the presence of particles.
For the sake of clarity, having made it clear that the operator is local, we omit the indices $x,y$ in the following notations. The coin operator thus results in the coin state $\ket{\circlearrowright}$ with probability $p_\text{a} = \Sigma_\text{a} \slash (\Sigma_\text{a} + \Sigma_\text{s})$ and it results in either of the four other states with probability $p_{s,i} = \Sigma_{\text{s},i} \slash (\Sigma_\text{a} + \Sigma_\text{s})$ respectively, where $i \in \mathcal{D} = \{\rightarrow, \uparrow, \leftarrow, \downarrow\}$. For simplicity, we consider that all scattering directions occur with the same probability $p_{\text{s},i} = p_\text{s} \slash 4$ with $p_{\text{s}} = \Sigma_\text{s} \slash (\Sigma_\text{a} + \Sigma_\text{s})$.

\subsubsection*{Diagonal unitary implementation}
\begin{figure}[H]
    \centering
    \includegraphics[width=0.95\linewidth]{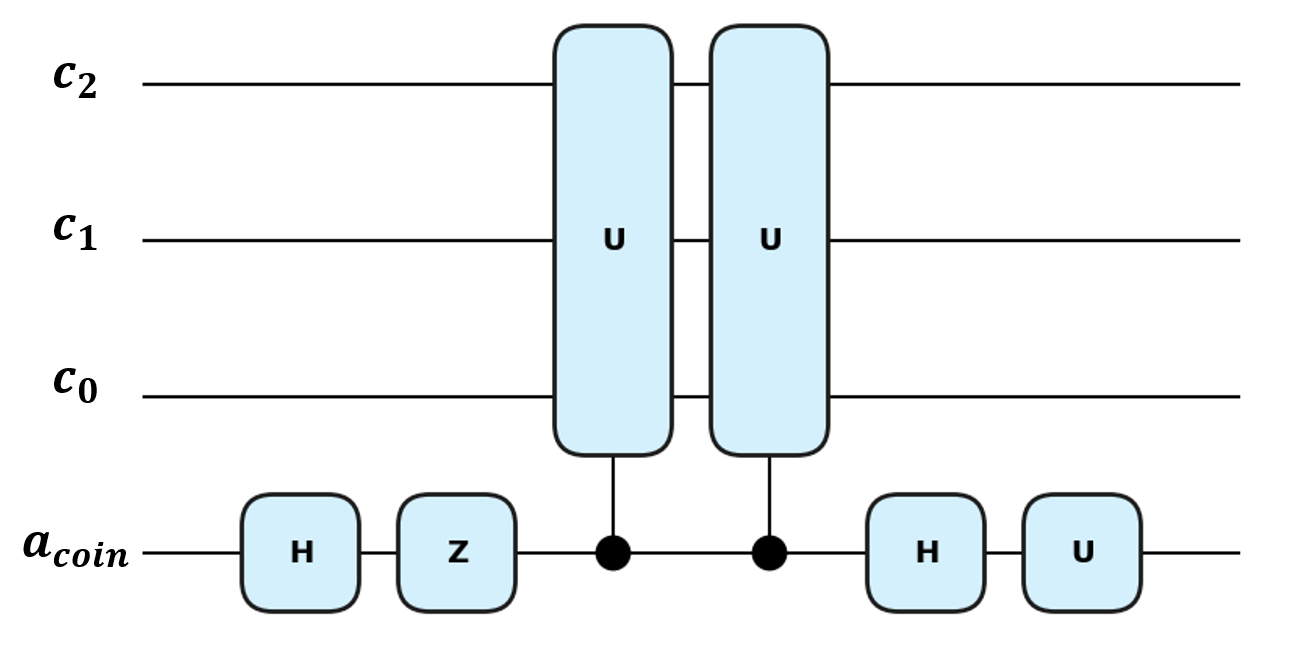}
    \caption{Non unitary diagonal operator}
    \label{fig:non_unitary_diagonal_operator}
\end{figure}

In order to implement a coin operator which is diagonal but not unitary, we rely on~\cite{zylberman2024efficient} whose circuit is depicted Figure~\ref{fig:non_unitary_diagonal_operator}.

The position-dependent coin implementation results in the construction of an operator represented by a matrix $C_{\ket{x,y}}$ of dimension $8 \times 8$ for a three-qubit coin register that is described by
\begin{equation*}
    C_{\ket{x,y}} =
    \left(\begin{array}{cc}
    {M_A}_{\ket{x,y}} & 0 \\
    0 & {M_S}_{\ket{x,y}}
\end{array}\right)
\end{equation*}
with
\begin{align*}
    {M_A}_{\ket{x,y}} &= \text{diag}\left(\frac{p_\text{a}^{x,y}}{4}, \frac{p_\text{a}^{x,y}}{4}, \frac{p_\text{a}^{x,y}}{4}, \frac{p_\text{a}^{x,y}}{4} \right) \\
    {M_S}_{\ket{x,y}} &= \text{diag}\left(p_{\text{s},\rightarrow}^{x,y}, p_{\text{s},\leftarrow}^{x,y}, p_{\text{s},\uparrow}^{x,y}, p_{\text{s},\downarrow}^{x,y} \right)
\end{align*}

\subsubsection*{Reflective boundary conditions}

\begin{figure}[H]
    \centering
    \includegraphics[width=0.95\linewidth]{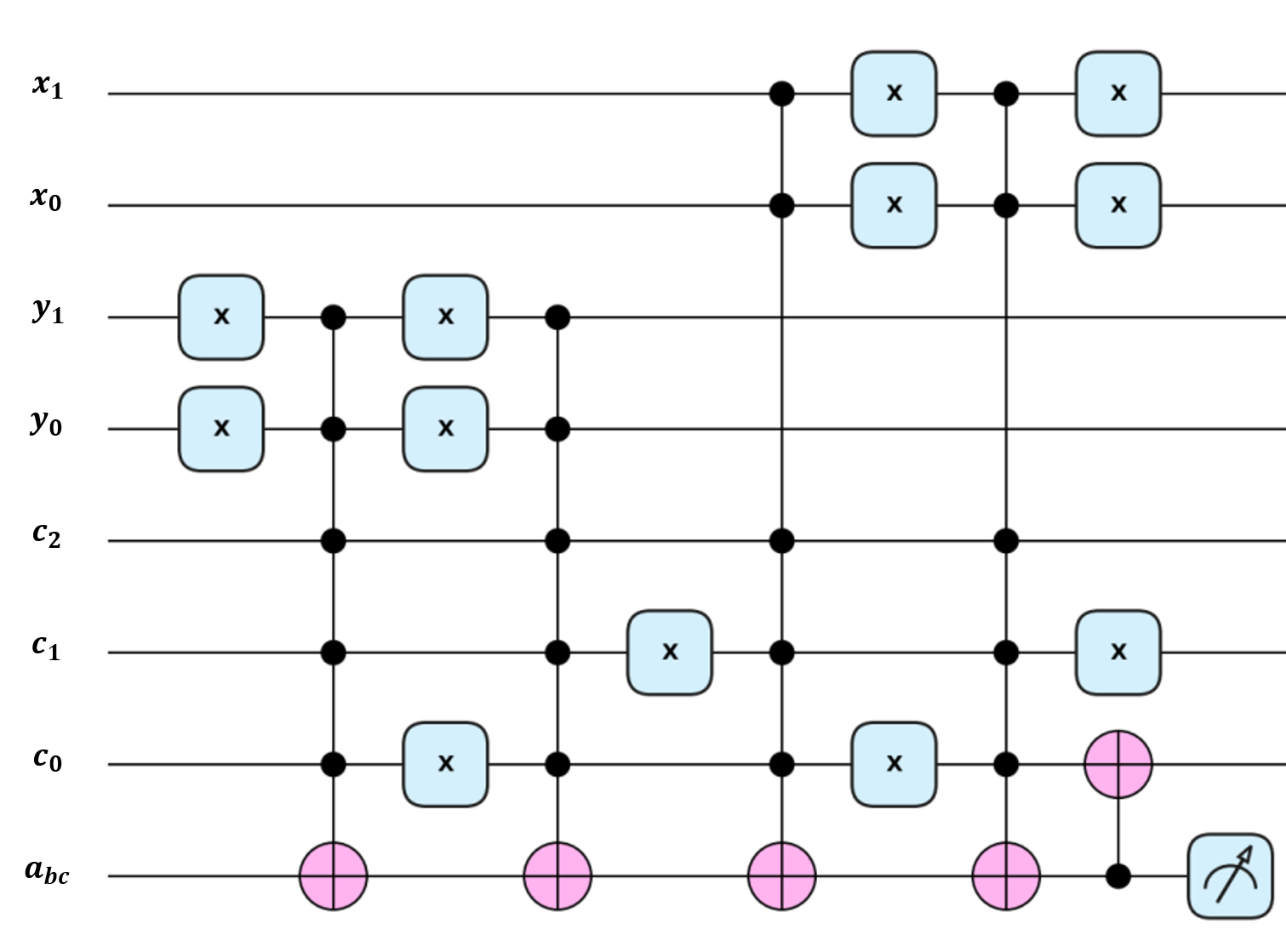}
    \caption{Circuit for the reflective boundary conditions.}
    \label{fig:reflective_boundary_conditions}
\end{figure}

Boundary conditions must be implemented when propagating the particles across the finite 2D geometry. In this work, they are designed such that the coin state is modified when it results in directions that are prohibited by the geometry. We make the choice of applying reflective boundary conditions, since periodic ones would not make physical sense for the problem studied.
The circuit proposed in Fig~\ref{fig:reflective_boundary_conditions} uses one ancillary qubit to be reset after each iteration.

Prohibited directions of propagation exist when the position state corresponds to a node on the edge of the geometry. This is the case when all qubits from either of the $x$- or $y$-register are in the state $\ket{1}$ (corresponding to the right and top edges of the geometry) or $\ket{0}$ (left and bottom edges of the geometry).
Therefore, the coin state is modified by flipping the coin qubit $\ket{c_0}$ (see previous Table) when the position register and coin register both combine into a prohibited movement.
This scheme is based on the global boundary conditions scheme presented in~\cite{koch2022gate}. The conditions operate on the set of states that compose each of the vertical and horizontal boundaries in our geometry. As opposed to the implementation proposed in~\cite{koch2022gate} where the coin operator is uncomputed before defining a new direction, our implementation only modifies the coin state by applying a CNOT gate at the cost of an additional ancillary qubit.
Another possible implementation of the boundary conditions could be made by incorporating them in the position-dependent coin.

\subsubsection*{Shift operator}

\begin{figure}[H]
    \centering
    \includegraphics[width=0.95\linewidth]{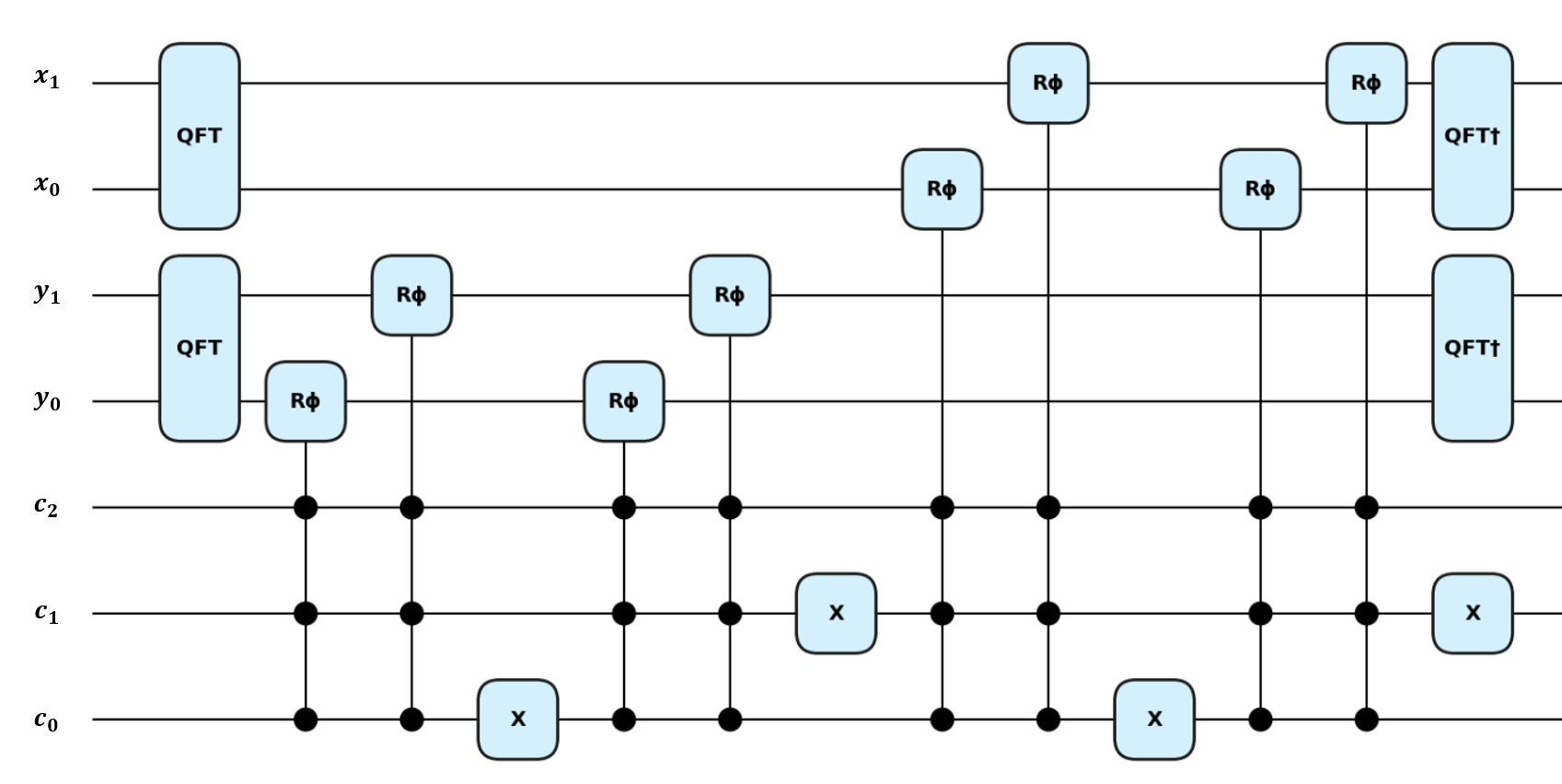}
    \caption{Circuit for the QFT-based shift operator.}
    \label{fig:shift_adder}
\end{figure}

The shift operator enables the transport of the particle by modifying the state of the position register based on the state of the coin qubits once the boundary conditions have been applied.
The chosen implementation of the shift operator is based on the quantum Fourier transform (QFT) as proposed in~\cite{koch2022gate}. Depending on the coin state, the qubits of the position register are either incremented or decremented such that the particle moves one step (or one node) in one of the predefined directions.
The QFT subroutine is applied in order to implement a "quantum adder" operator. As shown in Fig~\ref{fig:shift_adder}, the modification of the value of the $x$-qubits or $y$-qubits is performed via a cascade of controlled Z-rotations on the corresponding qubits. Keeping in mind the chosen convention $\ket{x_{n_x - 1} \ldots x_1 x_0}$ where $x_0$ is the least significant bit, the controlled rotation that acts on the position qubit $x_i$ is of angle $\pm \frac{\pi}{2^{n_x - 1 - i}}$ for $i \in \llbracket 0, n_x -1 \rrbracket$. The "$+$" sign is assigned when the particle is supposed to move one step to the right ($\ket{c_2c_1c_0} = \ket{\rightarrow})$ while the "-" sign is chosen if it is supposed to move one step to the left. An identical protocol is used on the $y$-qubits for the up/down directions.
It is important to note that, since the third coin qubit $c_2$ is also a control qubit, the shift operator as presented in Fig~\ref{fig:shift_adder} does not change the state of the position qubits when the coin state corresponds to $\ket{\circlearrowright}$. This corresponds to the absorption condition described above.

\subsection{Scoring a response}

\begin{figure}[H]
    \centering
    \includegraphics[width=0.6\linewidth]{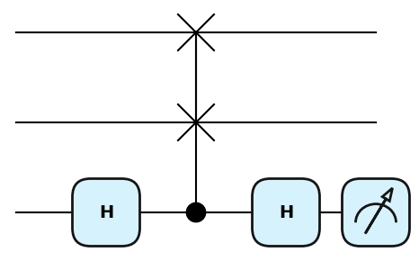}
    \caption{Swap test circuit}
    \label{fig:swap_test}
\end{figure}

\begin{figure*}[ht]
    \centering
    \includegraphics[width=0.99\linewidth]{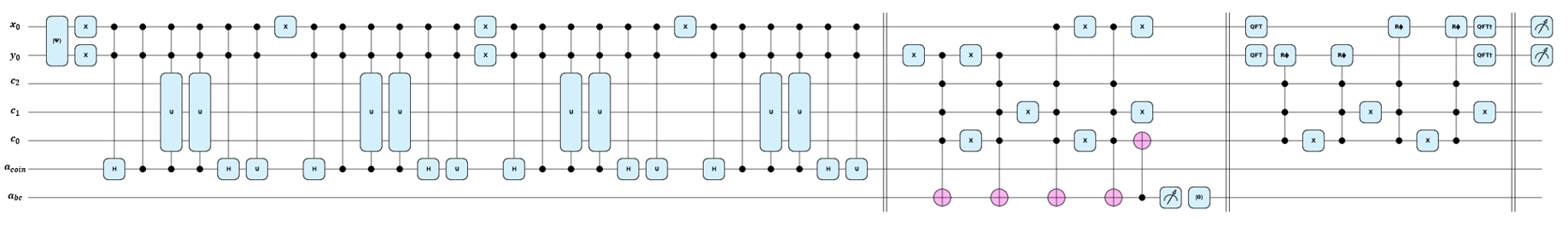}
    \caption{Circuit for one iteration and position sampling}
    \label{fig:particle_transport_circuit}
\end{figure*}

\begin{figure*}[ht]
    \centering
    \includegraphics[width=0.99\linewidth]{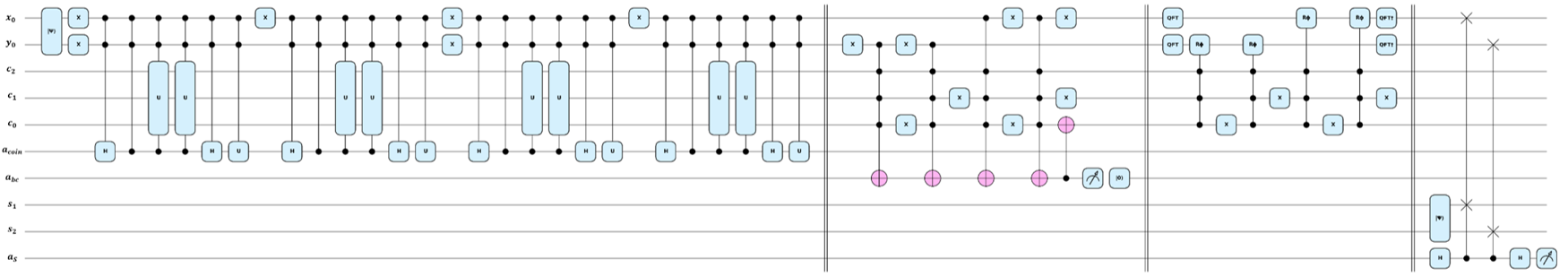}
    \caption{Circuit for one iteration and response score sampling}
    \label{fig:particle_transport_circuit_with_swap_test}
\end{figure*}

The state of the particle is encapsulated in the position register. After $N$ collisions, we can count the number of collisions on each node. If we wish to score a response locally, we can prepare a score register where the normalized prepared state represent the identity function on the region of interest and 0 elsewhere. By resorting to an ancilla qubit to measure the final integrated score, we apply the swap test on these position and score registers. See Figure~\ref{fig:swap_test}.
The final circuit is presented Figure~\ref{fig:particle_transport_circuit} if we wish to measure the positions directly.
The final circuit with the swap test against the scoring response is presented in Figure~\ref{fig:particle_transport_circuit_with_swap_test}.

\subsection{Measurement strategies}
Now that our circuits are defined, we can move on to the definition of the different measurement strategies.

\subsubsection*{Measured quantum walk}
The first strategy that we explore consists in measuring the position register at each time step~\cite{brun2003quantum, didi2022measurement}.
We measure $1000$ times the final probability distribution on the position register.
We then use these shots as the source for our next step.
The behaviour of our walk should converge to a classical one.

\subsubsection*{Amplified quantum walk}
The second strategy is to propagate the quantum walk with the coin, boundary conditions and shift operators $N$ times and then apply amplitude amplification~\cite{brassard2002quantum}.

\section{Numerical Experiments}
\label{sec:numerical_experiments}
\begin{figure*}[ht]
    \centering
    \subfloat[\label{fig:fd}Finite Differences]{\includegraphics[width=0.3\linewidth]{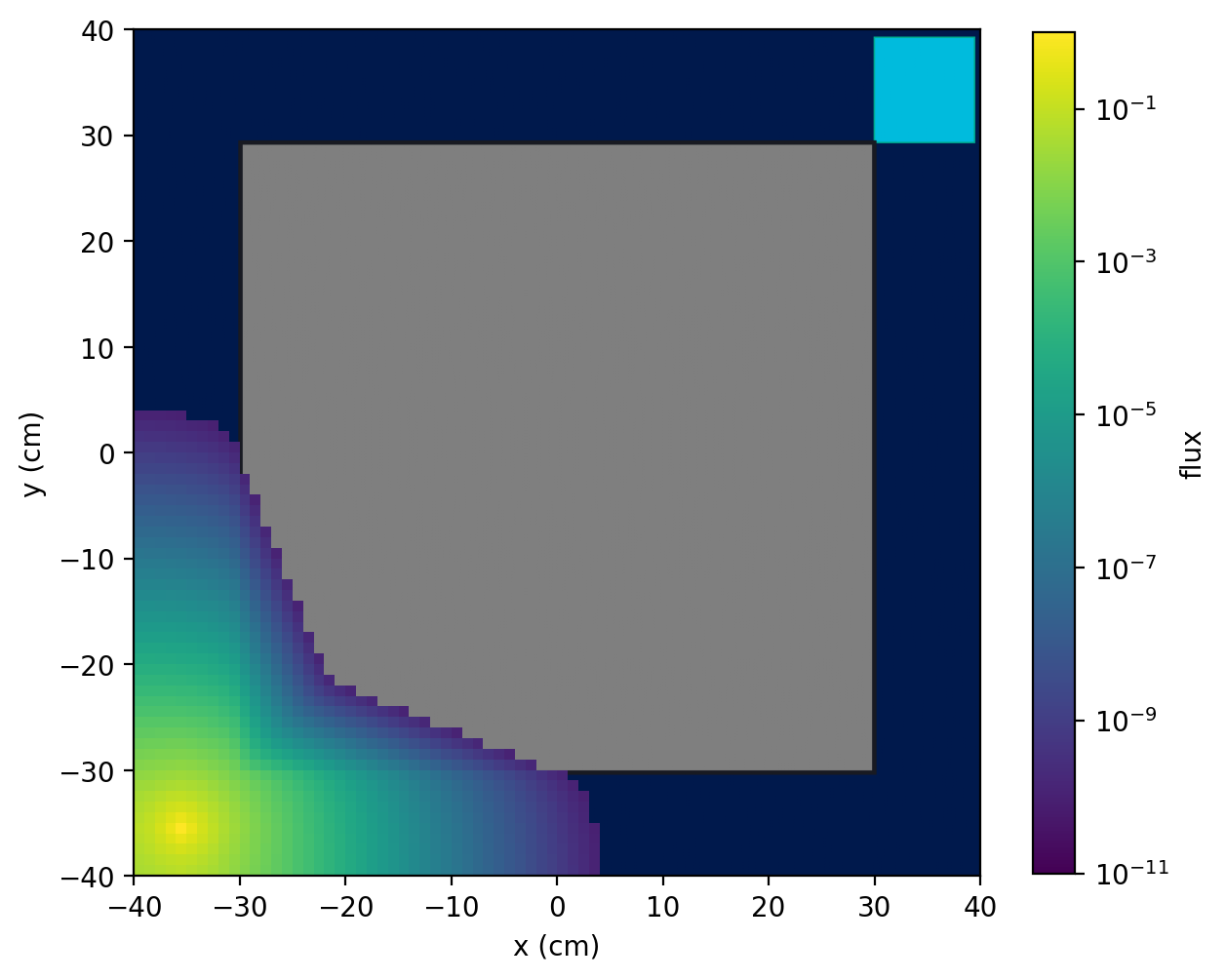}}
    \subfloat[\label{fig:mc}Monte Carlo]{\includegraphics[width=0.3\linewidth]{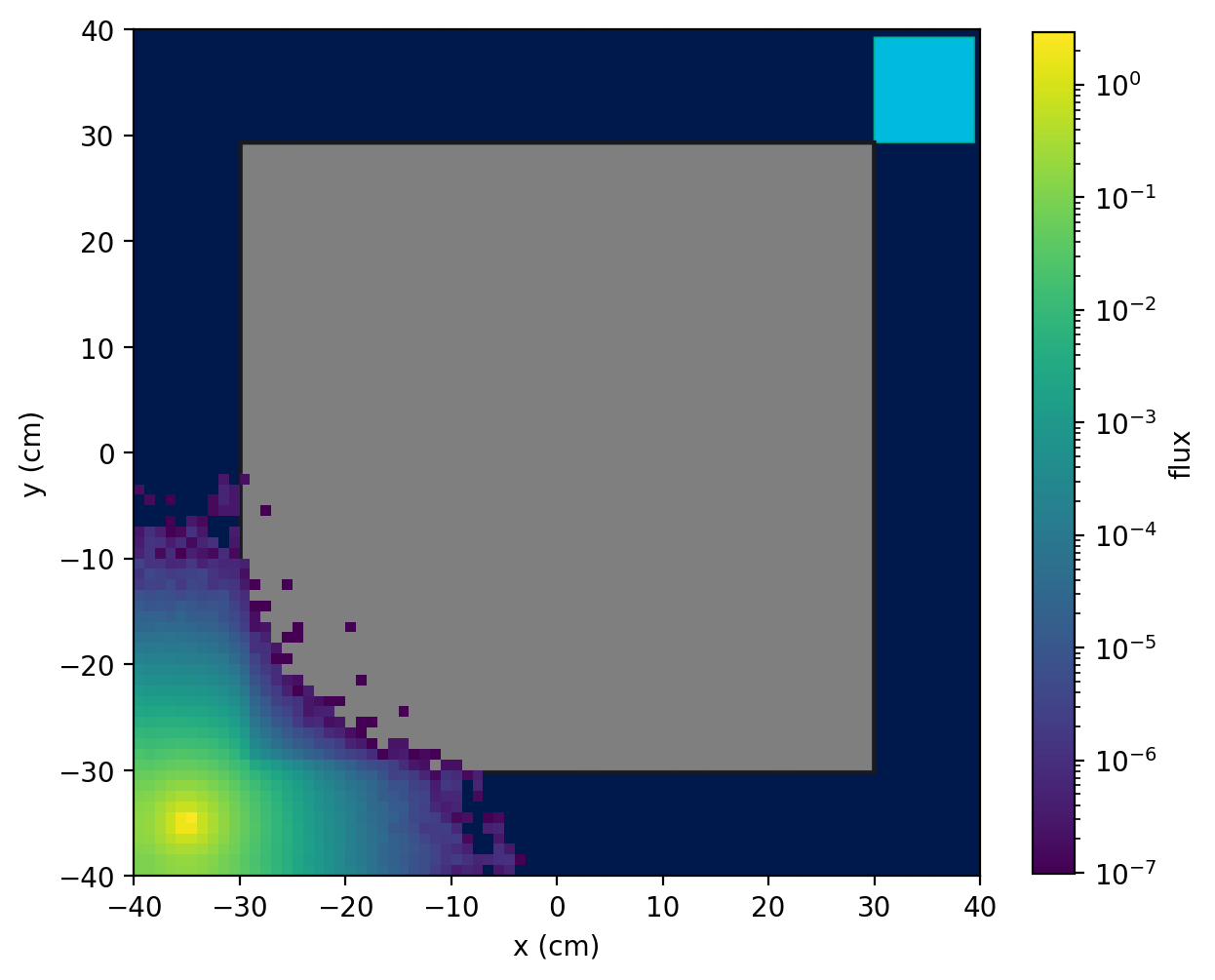}}
    \subfloat[\label{fig:mqw}measured quantum walk]{\includegraphics[width=0.3\linewidth]{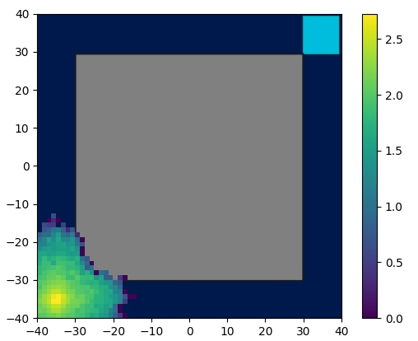}}
    \caption{Flux comparison}
    \label{fig:direct_fd}
\end{figure*}

\begin{figure}[H]
    \centering
    \includegraphics[width=\linewidth]{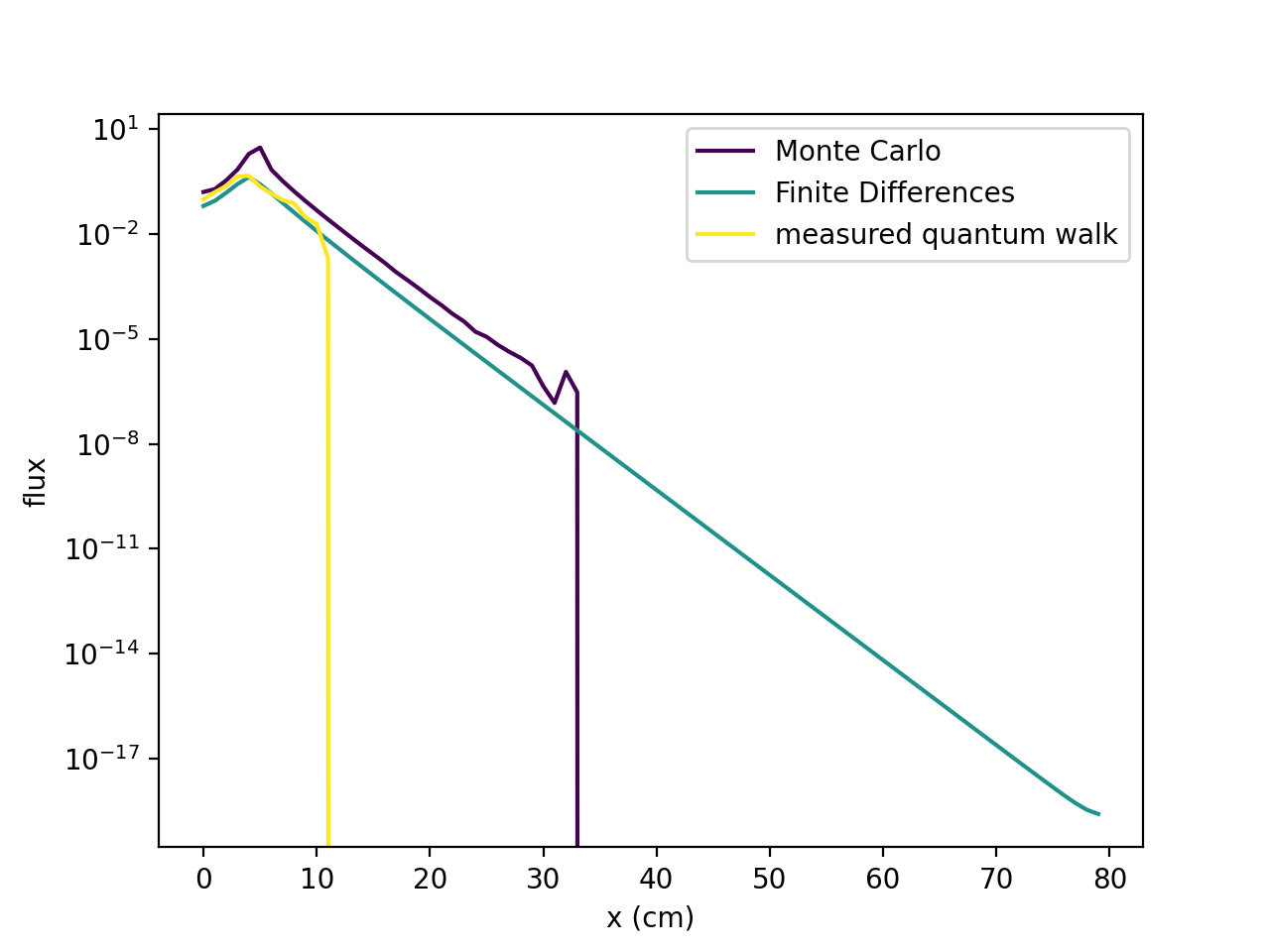}
    \caption{Comparison between classical Monte Carlo, classical Finite Differences and measured quantum walk.}
    \label{fig:comparison}
\end{figure}

This section is dedicated to numerical experiments.
We score the flux for each of our configurations.
Figure~\ref{fig:fd} shows the result for the classical Finite Differences scheme cropped at $10^{-10}$ for the flux.
Figure~\ref{fig:mc} shows the result for the classical Monte Carlo scheme with $500k$ particles.
Figure~\ref{fig:mqw} shows the result for the quantum random walk measured at each step.
We notice that the position dependent coin operator acts well and the geometry is well represented in the measured quantum walk plot.
In order to better compare the solutions, we slice the geometry at $x=5cm$ and plot the 1D flux in Figure~\ref{fig:comparison}.
We notice that the classical results agree within a normalizing constant.
For the measured quantum walk after 10 iterations, the flux qualitatively agrees but start diverging after tens of iterations. This is probably due to the fact that the absorption is not correctly taken into account.

\section{Conclusion}
In this work, we investigated the potential encoding of classical trajectories on a quantum computer and explored how we can define a score response.
We had to account for the fact that probabilities of interactions are local, thus resorting to a position dependent coin.
In order to apply the probabilities of interaction locally, we had to resort to non unitary tricks, which added an ancilla qubit to our final circuit.
Also, we discussed two ways of implementing reflective boundary conditions and implemented one of them.
This implementation relies on the partial measure of one qubit during the calculation, which increased the computational time significantly during our emulations.
We investigated two strategies for the propagation.
The first one relies on measurements at every time step, and on a reintegration of these shots as a source for the next step.
This strategy yields results that can be compared qualitatively to the classical deterministic and Monte Carlo solution obtained.
The second strategy consists in propagating the shift operator $N$ times before applying amplitude amplification.
The second strategy is computationally expensive as the boundary condition operator needs to be reset in the middle of the calculation.

Next research tracks for the future are the following.
First, our experiments were run on a simple 2D geometry, where materials have the same total cross sections $\Sigma_t=1$.
We only played with the scattering and absorption probabilities.
A perspective is thus to integrate properly the total cross section into numerical scheme.
Then, the directions of propagation were only left right up down. We can improve the reach of the particles by increasing the coin dimension and adapting the shift operator accordingly.
Finally, the fact that we consider a loop movement for the absorption is clearly an approximation that leads to discrepancies in the flux computation.
We should think of an alternative to remove the absorption from the collision probability density.

\bibliographystyle{unsrt}
\bibliography{main}

\end{multicols}
\end{document}